\documentstyle[twocolumn,ifthen,floats,aps,prl,psfig]{revtex}

\begin{document}
\draft

\twocolumn[\hsize\textwidth\columnwidth\hsize\csname@twocolumnfalse\endcsname%


\title{Spin transitions in a small Si quantum dot}

\author{L.P. Rokhinson, L.J. Guo\cite{guoaddr}, S.Y. Chou and D.C. Tsui}

\address{Department of Electrical Engineering, Princeton University,
Princeton, NJ 08544}

\date{\today}

\maketitle

\begin{abstract}
We have studied the magnetic field dependence of the ground state
energies in a small Si quantum dot.  At low fields the first five
electrons are added in a spin-up -- spin-down sequence minimizing the
total spin. This sequence does not hold for larger number of
electrons in the dot. At high fields the dot undergoes transitions
between states with different spins driven entirely by Zeeman energy.
We identify some features that can be attributed to transitions
between different spin configurations preserving the total spin of
the dot. For a few peaks we observed large linear shifts that
correspond to the change of the spin of the dot by 3/2. Such a change
requires that an electron in the dot flips its spin during every
tunneling event.
\end{abstract}


\vskip2pc]

The spin degree of freedom is an essential part of mesoscopic
physics. Quite often the knowledge of the spin advances our
understanding of electron-electron interactions, which can reveal
themselves in a non-trivial spin configuration of a mesoscopic
system. From this perspective, quantum dots can be regarded as model
systems for the study of spin--related phenomena because they contain
just a few electrons and different coupling parameters can be tuned
almost independently\cite{dotbook}. Within the simplest model of
non-interacting electrons each additional electron is added into the
dot to the next single-particle energy level (there is also a
constant energy associated with the charging of the environment). The
spin is accounted for by allowing two electrons to fill the same
single-particle energy level, thus the total spin of the system
should alternate between $s=0$ and $s=1/2$. There are several ways to
determine the spin of the dot. The most direct way is by studying the
Kondo effect\cite{goldhaber98a,cronenwett98,schmid98}. If the spin of
the ground state $s>0$, the Coulomb blockade is lifted in the
corresponding conduction valley at low temperatures. Indeed, valleys
with Kondo-enhanced conductivity were found to alternate with the
regular Coulomb blockade valleys of vanishing conductivity. However,
if the dot is weakly coupled to the leads the Kondo temperature can
be too small to be achieved experimentally. In such dots individual
energy levels are sharp and spin of the tunneling electron can be
determined from the Zeeman shift of the energy level.   The shift of
consecutive peaks has been shown to alternate with $\pm g^*\mu_BB$ in
small Al clusters\cite{ralph97} and carbon nanotubes\cite{cobden98},
supporting the alternating spin filling of the dot (here $g^*$ is the
effective $g$-factor and $\mu_B$ is the Bohr magneton). In lateral
semiconductor quantum dots with a large number of electrons the
Zeeman shift is masked by much larger orbital effects, and direct
determination of the spin is a formidable task.  Peaks fluctuate as a
function of $B$, reflecting the orbital shift of the levels. An
indirect information about the spin can be obtained from the
comparison of such "magnetic fingerprints", in order to find whether
the two consecutive electrons fill the same energy level. The
underlying assumption here is that the addition of an electron does
not change the spectrum of the dot significantly. Using this method,
strong deviations from the alternating spin filling has been
reported\cite{stewart97}. The most dramatic example of a
non-alternating spin filling is the polarization of small vertical
dots due to exchange interactions, similar to the Hund's rule in
atomic physics\cite{tarucha96}.

\begin{figure}[tb]
\psfig{file=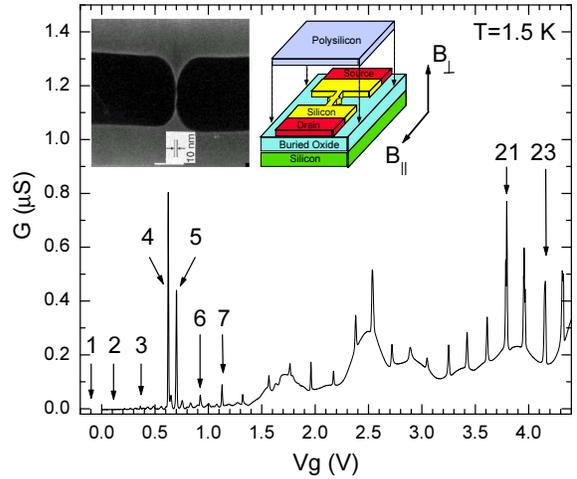,height=2.5in}\vspace{0.1in}
\caption{Conductance as a function of gate voltage measured with
$V_{ac}=100$ $\mu$V at $T=1.5$ K and $B=0$. Peaks are numbered in
sequence starting from the entrance of the first electron into the
dot (peak 1). The first two peaks are not resolved at zero bias but
their positions can be determined from the high-bias spectroscopy. In
the inset a schematic and a scanning electron micrograph of the
sample are shown.}
\label{g-vg}
\end{figure}

\begin{figure}[tb]
\psfig{file=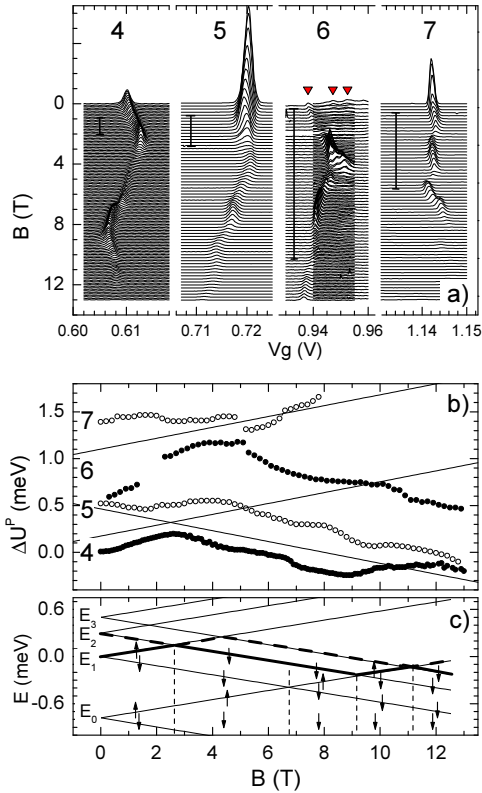,height=4.1in}\vspace{0.1in}
\caption{a) Evolution of four consecutive peaks as a function of $B
|| I$. Conductance was measured at 200 mK with $V_{ac}=50$ $\mu$V.
Individual traces are offset linearly with $B$ and bars are 1 $\mu$S
scales. In b) peak shifts $\Delta U^p(B)=[V_g^p(B)-V_g^p(0)]/\alpha$
are plotted for the same four peaks. The zero-field positions are
arbitrarily offset. Points are omitted if peak conductance $<0.01$
$\mu$S. Peak 6 is comprised of three peaks at $B<2$ T [marked with
triangles in a)] and only the lowest-energy branch is shown. Solid
lines have a slope of 0.058 meV/T ($1/2g^*\mu_B$ for $g^*=2$). c)
Schematic evolution of single-particle energy levels,
assuming that $B$-dependence enters only through the Zeeman energy.
Thin lines have slopes $\pm1/2g^*\mu_BB$. Thick solid and dashed
lines follow energies of the 4-th and 5-th electrons. Spins of the
four lowest states are indicated by small arrows.}
\label{lowpeaks}
\end{figure}

\begin{figure}[tb]
\psfig{file=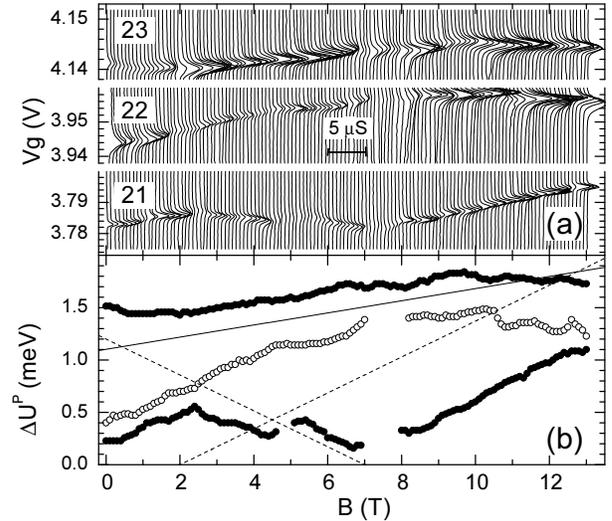,width=3.25in}
\caption{a) Evolution of peaks 21, 22 and 23 as a function of $B \bot
I$. Conductance was measured at $T=60$ mK using $V_{dc}=20$ $\mu$V.
All three data sets have the same scale. In b) peak shifts are
plotted for the same three peaks, similar to Fig.~\ref{lowpeaks}.
Solid and dashed lines have slopes 1/2 and 3/2$g^*\mu_B$
respectively.}
\label{highpeaks}
\end{figure}

The application of a magnetic field alters the spin configuration.
Energy levels shift differently with magnetic field and cross
each other. In the vicinity of such a magnetically-induced level
crossing exchange interactions may lift the degeneracy by favoring
the formation of a triplet state\cite{wiel98}.  A sufficiently large
field gradually polarizes the dot by collapsing all electrons into
the lowest Landau level\cite{mceuen91,ashoori96}.

In this Letter we examine the electron transport through a small Si
quantum dot.  Unlike the previously studied semiconductor dots, in
our dot the field-dependent shift of energy levels is dominated by
the Zeeman energy rather than by orbital effects.  Thus, we can
measure the spin of the dot directly for ground states with different
number of electrons, starting from one. We also study the evolution
of the total spin as a function of the magnetic field. We find that
the field dependence of energy levels consists of several linear
segments with different slopes. Comparison with a simple model for
non-interacting electrons allows us to identify a set of spin
transitions for the first few ground states. In addition to the
singlet-triplet and triplet-polarized transitions there are some
features in the spectra, which we attribute to transitions between
different realizations of the triplet state.  A detailed analysis
reveals some deviation from the model that hints for the importance
of the underlying interactions. As the number of electrons increases,
the spectra become more complicated. For example, tunneling of an
electron into the dot can change the total spin of the other
electrons in the dot. Such a tunneling process is beyond the scope of
a model of non-interacting electrons.


The measurements were performed on a small Si quantum dot fabricated
from a silicon-on-insulator wafer. The dot resides inside a narrow
bridge patterned from the top Si layer (see inset in
Fig.~\ref{g-vg}). A 50 nm thick layer of thermal oxide is grown
around the bridge followed by a poly-Si gate. The fabrication steps
have been described previously\cite{leobandung95}.  Gate capacitance
is estimated to be 0.8-1.0 aF, the total capacitance $C \approx15$ aF
and the  charging energy $U_c=e/C\approx 10$ meV. Spacing between
excited levels $\delta\sim 1-4$ meV, measured using non-zero bias
spectroscopy, is comparable to the charging energy and is consistent
with the lithographical size of the dot
$l\approx\sqrt{\hbar/m^*\delta}\approx100-190$ \AA\ . The gate
voltage -- to -- energy conversion coefficient, measured from both
non-zero bias spectroscopy and $T$-dependent scaling of the peak
width, is $\alpha\approx14$ mV/meV. The sample was studied in three
separate cooldowns and the reported phenomena were found to be
insensitive to redistribution of background charges, thus reflecting
intrinsic properties of the dot.

At high temperatures, $T>120$ K, the device exhibits regular
metal-oxide-semiconductor field effect transistor (MOSFET)
characteristics with a threshold gate voltage of $V_{th}\approx-0.2$
V. At $T<100$ K the Coulomb blockade emerges and the conductance
oscillates as a function of the gate voltage $V_g$ for $V_g>V_{th}$.
A representative trace of the conductance $G$ as a function of $V_g$
at $T=1.5$ K is shown in Fig.~\ref{g-vg}. There is a series of sharp
peaks, spaced by 150-200 mV. The peaks, corresponding to the entrance
of the first two electrons, cannot be reliably measured, but their
positions can be determined from high bias spectroscopy. Commonly for
this type of devices, the sample has a parallel conducting
channel\cite{rokhinson00a}, which exhibits Coulomb blockade at
$V_g<1.2$ V with peaks separated by 60 mV. At $V_g>1.2$ V the extra
channel has finite conductance with some broad features as a function
of $V_g$. Coulomb blockade peaks, originated from the lithographical
dot, are not broadened at high $V_g$, thus electrical transport
through the dot and the parallel channel are decoupled. Charging of
remote impurities (which can be accomplished by wide gate voltage
scans $\Delta V_g>3$ V) changes positions of the extra peaks at
$V_g<1.2$ V and the value of the background conductance at $V_g>1.2$
V without altering the position and amplitude of the main peaks.

We studied the peak positions $V_g^p$ as a function of magnetic field
$B$ for the first 30 peaks. For $V_g<0.4$ V (the first three peaks),
electron density in the contacts is low and the contacts are spin
polarized by a moderate magnetic field. For $V_g>0.4$ V ($N\geq4$)
both spin subbands in the contacts are occupied within the
experimental range of $0<B<13$ T. Thus, the Fermi energy $E_F$ is
field-independent and the peak shift reflects only the field
dependence of the energy levels in the dot (mobility of the
two-dimensional gas is low, $\approx 300$ cm$^2$/V$\cdot$s at 4.2 K,
and there is no measurable modulation of $E_F$ due to
Subnikov-de-Haas oscillations for $B$ up to 13 T).


The evolution of several peaks with $B$ is shown in Fig.
\ref{lowpeaks}a (peaks 4-7) and in Fig.~\ref{highpeaks}a (peaks
21-23). Clearly, $V_g^p$ and the peak amplitudes $G^p$ change
non-monotonically with $B$. Analysis of "magnetic fingerprints"
reveals that there is no apparent pairing of the neighboring peaks
within the first 30 peaks. In fact, we observed an unexpected
tripling of the peaks: two bunches of peaks have similar "magnetic
fingerprints" for three consecutive peaks (13,14,15 and 16,17,18, not
shown in the figures). The measurements were repeated for two
different orientations of $B$, defined in the inset in
Fig.~\ref{g-vg}. We found that $V_g^p$ is insensitive to the
direction of the magnetic field: aligning $B$ with the current
direction ($B_{||}$, in-plane) or perpendicular to the plane of the
sample ($B_\bot$) does not change $V_g^p$ significantly. The dot is
lithographically asymmetric and the orbital effects are expected to
depend on the field direction. Thus, we conclude that in our small
dot the $B$-dependence of $V_g^p$ is dominated by spin effects. This
conclusion is also supported by the observation that, in the range of
$B$ when the contacts are fully spin-polarized, the $V_g^p$ for peaks
1-3 does not depend on $B$ at all.

Unlike $V_g^p$, the peak amplitude $G^p$ depends on the direction of
the magnetic field. The $G^p$ reflects the tunneling probability and
depends exponentially on the overlap of wavefunctions in the dot
and in the contacts.  As such, $G^p$ is sensitive to a particular
configuration of the wavefunction within the dot, and redistribution
of the wavefunction due to small orbital effects can result in a
significant change of $G^p$.

What physics is behind the field-dependence of the peak position? At
zero bias $V^p_g$ is determined by the degeneracy condition that the
electrochemical potentials for the ground states with $N-1$ and $N$
electrons in the dot are equal. Provided that the Fermi energy in the
contacts is independent of the magnetic field, a shift of the $N$-th
peak with $B$ reflects the relative change of the ground state
energies $\Delta U^p_N(B)=\Delta U(N,B)-\Delta U(N-1,B)$, where
$U(N,B)$ is the energy of the ground state of $N$ electrons in
magnetic field $B$ and  $\Delta U(N,B)=U(N,B)-U(N,0)$. In the absence
of spin-orbit interactions (which is the case for the bulk Si) the
total energy can be separated into spin and orbital terms. The spin
term includes Zeeman energy $s(N)g^*\mu_BB$ where $s(N)$ is the total
spin of the ground state with $N$ electrons and $g^*$ is the
effective $g$-factor. Thus, the Zeeman-related peak shift is
$[s(N)-s(N-1)]g^*\mu_BB=(\pm n\pm1/2)g^*\mu_BB$, where the spin
$\pm1/2$ is carried by the tunneling electron and $n=0, 1, 2, \ldots$
is the number of electrons in the dot that flip their spins upon the
tunneling event.  In the simplest case of no interactions peaks
should shift linearly with $B$ by $\pm1/2g^*\mu_BB$.

Experimentally, peaks do not shift linearly with $B$. Instead,
$dV^p_g/dB$ changes both its value and sign as $B$ is varied from 0
to 13 T. For a quantitative analysis, peak positions are extracted
from $G$ vs $V_g$ scans, and the peak shifts $\Delta
U^p(B)=[V_g^p(B)-V_g^p(0)]/\alpha$ are plotted as a function of $B$
in Figs.~\ref{lowpeaks}b and \ref{highpeaks}b. The curves are offset
for clarity. For a comparison, lines with slopes $\pm1/2g^*\mu_B$ for
$g^*=2$ are also shown (solid lines). First, let us focus on the
low-field ($B<2$ T) region.  Peaks 4 and 5 shift linearly with $B$
and the corresponding slopes are + and $- 1/2g^*\mu_B$. In the same
low-field region the preceding peaks 2 and 3 also shift with + and $-
1/2g^*\mu_B$ slopes correspondingly. Thus, at low fields the ground
states with up to 5 electrons in the dot have the lowest spin
configuration and the dot is filled in a spin-down -- spin-up
sequence $\downarrow\uparrow\downarrow\uparrow\downarrow$ (in the
order the levels are filled). Such a filling sequence requires that
the valley degeneracy is lifted and two electrons with different
spins can occupy the same energy level.

This simple picture of alternating filling does not hold for $N>5$
even at low fields. At $B<2$ T peak 6 consists of three peaks
separated by $\approx0.5$ meV at zero field, none of which shifts
with $1/2g^*\mu_BB$ (the zero-field positions of the three peaks are
marked by triangles in Fig.~\ref{lowpeaks}a). The slope of the
lowest-energy branch is close to $3/2g^*\mu_B$, the other two
branches have small negative slopes. The shift of the next, the 7-th,
peak has a positive slope, while the lowest-spin arrangement for a
dot with 7 electrons should have negative Zeeman energy.  We conclude
that the ground state with 6 electrons is spontaneously polarized and
the total spin $s(6)>1/2$. Transitions between ground states that
involve a change in spin by $\Delta s>1/2$ have low probability and
the corresponding peaks are expected to be suppressed (so-called spin
blockade\cite{weinman95}). Indeed, the overall conductance of peak 6
is strongly suppressed and, presumably, the appearance of several
branches can be explained by the instability of the polarized state.

The low-field spin configuration is not preserved at high magnetic
fields.  For peak 4, $dV^p_g/dB$ changes sign from positive to
negative at $B=2.5$ T, back to positive at $B=9$ T, and, again, to
negative at $B\approx12$ T.  The spin of the tunneling electron
changes from being
$+1/2\rightarrow-1/2\rightarrow+1/2\rightarrow-1/2$. The
corresponding spin transitions of the ground state can be understood
from a simple model for non-interacting electrons.  Let us consider
four single-particle levels $E_i$, as shown in Fig.~\ref{lowpeaks}c.
Each level is spin-degenerate at zero field and splits into two
levels $E_i\pm1/2g^*\mu_BB$ for $B>0$.  In the absence of
interactions position of the $N$-th peak is determined by
$U(N,B)-U(N-1,B)=\sum^N_k{E(k,B)}-\sum^{N-1}_k{E(k,B)}=E(N,B)$, where
$E(k,B)$ is the energy of the $k$-th electron, including the Zeeman
contribution.  $E(4,B)$ is outlined by the thick solid line in Fig.
\ref{lowpeaks}c.  Qualitatively, $E(4,B)$ captures the main features
of $V^p_g$ vs. $B$ for the 4-th peak and the kinks can be attributed
to the corresponding level crossings.  Each level crossing results in
a change of the spin configuration and the ground state of 4
electrons undergoes spin transitions as a function of $B$:
$\downarrow\uparrow\downarrow\uparrow\rightarrow
\downarrow\uparrow\downarrow\downarrow\rightarrow
\downarrow\downarrow\uparrow\downarrow\rightarrow
\downarrow\downarrow\downarrow\uparrow\rightarrow
\downarrow\downarrow\downarrow\downarrow$ (the regions with different
spin configurations are separated by dashed vertical lines in Fig.
\ref{lowpeaks}c). The first transition is singlet-triplet and the
last transition is triplet-spin polarized. There are two intermediate
transitions within the triplet state which change the spin
configuration within the dot without changing the total spin. At
$B\approx7$ T the spin configuration of the ground state with 4
electrons changes without reversing the spin of the tunneling
electron; such a transition does not change the sign of $dV_g^p/dB$.
In the absence of interactions there should be no corresponding kink.
The second transition flips the spin of the tunneling electron and of
an electron in the dot simultaneously, preserving $s(4)=-1$ but
changing the sign of $dV_g^p/dB$.

The model, described above, also reproduces the features of peak 5
for $B<7$ T (dashed line in Fig.~\ref{lowpeaks}c).  However, there
are some important discrepancies, which cannot be understood within
this model of non-interacting electrons.  First of all, we cannot
describe the evolution of $N>5$ peaks within this model.  Second,
each level crossing should result in a pair of upward--downward kinks
in two neighboring peaks at the same value of $B$.  Clearly, kinks in
$V_g^p(B)$ for peaks 4 and 5 near 2 T are shifted by $\approx0.5$ T.
The most notable deviation from this simple model of level crossing
is shown in Fig.~\ref{highpeaks}b, where upward kinks at 2.3 T and
5.3 T in $V_g^p(B)$ for peak 21 have no downward counterparts in
$V_g^p(B)$ for peak 22.  Third, we have to assume a small
single-particle level spacing of $\approx0.3$ meV to fit the
positions of the observed spin transitions.  From non-zero bias
spectroscopy, as well as from the statistics of the zero-bias peak
spacing, we estimate that excited levels are separated by 1-4 meV.

For most of the peaks $|dV^p_g/dB|\approx1/2g^*\mu_B$.  However,
there are a few peaks that shift much faster with magnetic field. In
Fig.~\ref{highpeaks}b $\Delta U^p(B)$ for peaks 21 and 22 have linear
segments with a slope $\approx3/2g^*\mu_BB$. Remarkably, the shift of
peak 21 has such a large slope in the whole range $0<B<13$ T,
although its sign changes four times.  We can rule out enhancement of
the $g$-factor because i) there are segments in the neighboring peak
23 with the slope $1/2g^*\mu_B$ (assuming $g^*=2$), and ii) it is
known that  interactions renormalize $g^*$ at low electron densities
in Si-MOSFETs but $g^*$ approaches the bulk value of 2 as the density
increases\cite{ando82}. Thus, the total spin of the dot changes by
$s(N)-s(N-1)=3/2$. A change of the spin by more than 1/2 means that
at least one electron in the dot should flip its spin
($3/2=1+1/2=2-1/2$) upon the tunneling of an electron. We want to
stress the difference with the spin transitions discussed above:
there, the total spin of the dot changes as a function of $B$, but it
is fixed for any particular $B$.  In order to change the total spin
by 3/2 an electron in the dot has to flip its spin during the
tunneling event.  In the absence of spin-orbit interactions such a
flip is forbidden unless some other spin scattering mechanism is
considered. As we mentioned earlier, the absence of an efficient spin
scattering should result in a spin blockade with the corresponding
suppression of the peak amplitude.  Experimentally, there is no
apparent suppression of peaks 21 and 22, which have the $3/2g^*\mu_B$
slopes, compared to the amplitude of peak 23, which has the regular
slope of $1/2g^*\mu_B$.

To summarize, we have analyzed the field dependence of ground state
energies in a small Si quantum dot.  The dot is in a new regime where
the $B$-dependence of the energy levels is dominated by the Zeeman
energy. There are distinctive features in the data which we attribute
to the transitions between different spin configurations of the dot.
For the state with 4 electrons in the dot we identified five
different spin configurations, including three with the same total
spin $s=-1$. Some peaks have large shift as a function of magnetic
field which requires the total spin of the dot to be changed by
$\Delta s>1/2$ upon the tunneling of an electron. Surprisingly, we
found that such peaks are not necessarily suppressed.

We gratefully acknowledge discussions with Boris Altshuler and
Richard Berkovits. The work was supported by ARO, ONR and DARPA.


\end{document}